\documentclass[journal]{IEEEtran}

\usepackage[english]{babel}
\usepackage[utf8]{inputenc}
\usepackage{amsmath}
\usepackage{amsthm}
\usepackage{amsfonts}
\usepackage{mathtools}
\usepackage{graphicx}
\usepackage{float}
\usepackage[justification=centering]{caption}
\usepackage{hyperref}
\usepackage{dsfont}
\usepackage{bbm}
\usepackage{multirow}
\usepackage{flushend}

\newtheorem{theorem}{Theorem}
\newtheorem{definition}{Definition}
\newtheorem{assumption}{Assumption}
\newtheorem{lemma}{Lemma}
\newtheorem{remark}{Remark}
\newtheorem{corollary}{Corollary}

\DeclareMathOperator*{\logdet}{log\,det}
\DeclareMathOperator*{\argmin}{arg\,min}
\DeclareMathOperator*{\tr}{\text{tr}}

\title{\LARGE \bf
Tube-based Guaranteed Cost Robust Model Predictive Control for Linear Systems  Subject to Parametric Uncertainties
}

\author{Carlos M. Massera$^{1,3}$, Marco H. Terra$^{2}$ and Denis F. Wolf$^{3}$
\thanks{$^{1}$Lyft Inc., 2300 26th St, San Francisco, CA, USA 94107 {\tt\small cmassera@lyft.com}}%
\thanks{$^{2}$São Carlos School of Engineering, University of São Paulo, Avenida Trabalhador São-carlense, 400, São Carlos, Brazil {\tt\small terra@sc.usp.br}}%
\thanks{$^{3}$Institute of Mathematics and Computer Science, University of São Paulo, Avenida Trabalhador São-carlense, 400, São Carlos, Brazil {\tt\small denis@icmc.usp.br}}%
}

\begin{document}
\maketitle
\thispagestyle{empty}
\pagestyle{empty}

\begin{abstract}
We propose a tube-based guaranteed cost model predictive controller considering a homothetic formulation for constrained linear systems subject to multiplicative structured norm-bounded uncertainties. 
It provides an upper bound to the general min-max model predictive control. The invariance property of the proposed tube holds for any arbitrary scaling. 
It yields in a second order cone programming problem which is less computationally expensive than standard 
semi-definite programming problems. 
We also present a numerical example with a comparative study among the proposed approach, an open loop guaranteed cost model predictive controller  and a homothetic tube model predictive controller for linear difference inclusion systems.
\end{abstract}


\section{INTRODUCTION}

Model predictive control (MPC) is a class of optimization-based control algorithms that use an explicit model of the controlled system to predict its future states \cite{badgwell2015model}. 
Several different fields have applications which use this technique, such as refineries, food processing plants, mining, aerospace, and automotive control \cite{qin2003survey}. 
An MPC minimizes a cost function while maintaining the system states and control inputs within a feasible set. 
The system dynamics are usually assumed to be known, and it does not consider model mismatches or external disturbances. However, the disregard for such uncertainties may lead to degraded closed-loop performance and violation of state and control input constraints \cite{rawlings1994nonlinear}.

Robust MPC (RMPC) approaches have been proposed to address MPC's poor closed-loop performance when the system is subject to uncertainties \cite{bemporad1999robust}. Scokaert and Mayne \cite{scokaert1998min} proposed a dynamic programming-based RMPC which provides an exact solution for bounded uncertainties. However, this approach yields an infinite dimensional optimization in the general case and exponential dimensional for polytopic bounded uncertainties. It results in intractable or prohibitively large optimization problems. Meanwhile, L{\"o}fberg \cite{lofberg2003approximations} proposed an approximate solution to RMPCs with $l_1$- and $l_{\infty}$-norms cost functions and polytopic-bounded additive disturbances. L{\"o}fberg \cite{lofberg2002minimax} also investigated approximate RMPC synthesis 
with $l_2$-norm bounded multiplicative disturbances. However, the proposed approach for such system relies on online solution of semi-definite programming (SDP) problems. It increases overall computational requirements in comparison to linear programming (LP), quadratic programming (QP) and second order coning programming (SOCP).

Tube model predictive control (TMPC) (\cite{mayne2005robust}, \cite{rakovic2012homothetic}, \cite{rakovic2012parameterized},
\cite{rakovic2016elastic}) is a subclass of RMPC which tries to ensure computational tractability while obtaining reasonably tight approximations of the exact solution. There are four categories of TMPC policies: rigid, homothetic, parameterized and elastic. Rigid TMPC (RTMPC) \cite{mayne2005robust} is the first proposed type of TMPC, where the tube is a translation of an off-line computed robust positive invariant set with a fixed cross-section. Homothetic TMPC (HTMPC) \cite{rakovic2012homothetic} is a direct extension of RTMPC, which parameterizes the tube by its translation and by scaling of its cross-section. Parameterized TMPC (PTMPC) \cite{rakovic2012parameterized} generalizes the concept of TMPC thought superposition and separability of additive disturbances. It allows for a  less conservative approximation of exact RMPC at the expense of increased computational cost. PTMPC scales quadratically with prediction horizon, while RTMPC and HTMPC scale linearly.  Elastic TMPC (ETMPC) \cite{rakovic2016elastic} is a generalization of HTMPC to reduce its conservativeness when compared to PTMPC while maintaining linear complexity scaling. The tubes in ETMPC are parametrized based on its translation and an arbitrary scaling for each of its facets.

According to Mayne et al. \cite{mayne2011robust}, one of the main limitations of TMPC is the lack of robustness to multiplicative disturbances. Evans et al. \cite{evans2012robust} proposed an extended TMPC for the case of stochastic multiplicative uncertainties. Meanwhile, Flemings et.al. \cite{fleming2015robust} proposed a RTMPC for polytopic-bounded multiplicative uncertainties. However, both of these methods only consider disturbances in the state transition matrix. Rakovi{\'c} and Cheng \cite{rakovic2013homothetic} proposed an HTMPC approach to deal with polytopic-bounded multiplicative uncertainties. 
Nevertheless, their proposed method is not able to completely isolate the uncertain error dynamics from the nominal system dynamics through the homothetic tube formulation.


In this paper, we propose a tube-based guaranteed cost model predictive controller (T-GCMPC) through a homothetic formulation. It extends the application of TMPCs to constrained linear systems subject to multiplicative structured norm-bounded uncertainties. It differs from similar homothetic TMPC formulations in three aspects:%
\begin{enumerate}
\item It provides a cost function associated with the non-robust synthesis counterpart. When disturbances are not present the cost reduces to the nominal MPC case.
\item It internalizes all uncertainties into the homothetic tube formulation. It guarantees that its cross-section volume tends to zero.
\item It introduces a novel invariant tube definition, where any arbitrary scaling of its cross-section is an invariant set for a subset of the state-space.
\end{enumerate}%
The first issue is guaranteed thanks to the GCC synthesis. To support the latter difference  we also propose a minimum robust control invariant level-set for this class of systems. The effectiveness of the T-GCMPC proposed is checked through a comparative study with the open loop (OL)-GCMPC of \cite{massera2017guaranteed} and the HTMPC for polytopic-bounded multiplicative uncertainties of \cite{rakovic2013homothetic}. We also present in the companion paper \cite {massera2019tubeaut}, an application of the T-GCMPC in an autonomous vehicle.

This paper is organized as follows: in Section \ref{sec_preliminaries} we present the preliminary results; in Section \ref{sec_mrci} we formulate the robust positive invariant level-sets and their application to this controller design; in Section we \ref{sec_tgcmpc} derive the tube-based guaranteed cost model predictive controller; in Section \ref{sec_examples} we provide a numerical example; finally, in Section \ref{sec_conclusion} we address the final remarks.

\section{PRELIMINARY RESULTS}
\label{sec_preliminaries}

We consider discrete-time linear system subject to parametric uncertainties of the form
\begin{equation}
\begin{aligned}
x_{k+1} &= (A + \delta A(\Delta_k)) x_k + (B^u + \delta B^u(\Delta_k)) u_k
\end{aligned}
\label{eq_sys_model}
\end{equation}%
where $ x_k \in \Re^{n_x} $ is the system state, $ u_k \in \Re^{n_u} $ is the control input, $ A \in \Re^{n_x \times n_x} $ is the state matrix, $ B^u \in \Re^{n_x \times n_u} $ is the input matrix, and $ \delta A(\Delta_k) $, $ \delta B^u(\Delta_k) $ are, respectively, the state and input multiplicative uncertainty matrices, such that%
\begin{equation}
\begin{bmatrix}
\delta A(\Delta_k) & \delta B^u(\Delta_k)
\end{bmatrix}
= 
B^w \Delta_k
\begin{bmatrix}
C_y & D_y^u
\end{bmatrix}
\label{eq_sys_uncertainty_matrix}
\end{equation}%
where $ B^w \in \Re^{n_x \times n_p} $, $ C_y \in \Re^{n_q \times n_x} $, $ D_y^u \in \Re^{n_q \times n_u} $, and $ \Delta_k \in \mathbb{D} $ where%
\begin{multline}
\mathbb{D} = \{ \Delta_k \mid \forall i \in [1, s]: \Delta_{k,i} \in \Re^{n_{pi} \times n_{qi}}, || \Delta_{k,i} ||_2 \le 1, \\ \Delta = \text{diag}(\Delta_1, \Delta_2, \ldots,\Delta_s) \}.
\label{eq_uncertainty_structure}
\end{multline}%

In order to evaluate the predictive approach we are dealing with in this paper, we rewrite System \eqref{eq_sys_model} in the following equivalent form
\begin{equation}
\begin{aligned}
x_{k+1} &= A x_k + B^u u_k + B^w w_k\\
y_{k} &= C_y x_k + D_y^u u_k\\
w_k &= \Delta_k y_k
\end{aligned}
\label{eq_sys_model_2}
\end{equation}%
where parametric uncertainties of \eqref{eq_sys_model} are considered as disturbances in a feedback system formulation.

\subsection{Optimal Min-Max Robust Model Predictive Control}

 Rigid, homothetic and elastic TMPC formulations have taken into account cost functions based on data from nominal models and tube relaxation variables. In this paper, we consider a cost function defined on states subject to disturbances. 
\begin{eqnarray}
J_i(x_i, \textbf{u}, N) = x_N^T P_N x_N +\nonumber  \\ \underset{k = i}{\overset{N-1}{\sum}} (x_k^T Q x_k + u_k^T R u_k + 2 x_k^T N u_k)
\label{eq_cost_function}
\end{eqnarray}
where $ P_N \succeq 0 $, $ Q \succeq 0 $, and $ R \succ 0 $ are symmetric weighting matrices, and $ \mathbf{u} = \{u_k \mid k \in [0, N-1]\} $.

In the scope of this paper, we are interested in investigating the T-GCMPC synthesis under the following assumptions:


\begin{assumption}
 For all $ \Delta_k \in \mathbb{D} $, the pair ($A + \delta A(\Delta_k)$, $B + \delta B(\Delta_k)$) is stabilizable.
\label{ass_stabilizable}
\end{assumption}

\begin{assumption}
For all $ \Delta_k \in \mathbb{D} $, the pair ($A + \delta A(\Delta_k)$, $Q^{\frac{1}{2}}$) is observable.
\label{ass_observable}
\end{assumption}

We are now ready to define the general infinite horizon min-max robust model predictive controller.

\begin{definition}
Consider the linear system model from \eqref{eq_sys_model}. Then, the (closed loop) infinite horizon min-max robust model predictive control is given by%
\begin{equation}
\begin{matrix*}[l]
J^*(x_0) = & \underset{u_0}{\inf} \; \underset{\mathbf{\Delta_0}}{\sup} \; \underset{u_1}{\inf} \; \underset{\mathbf{\Delta_1}}{\sup} \dots \underset{N \rightarrow \infty}{\lim} J_0(x_0, \mathbf{u}, N)\\
& s.t. \: x_{k+1} = A x_k + B^u u_k + B^w w_k\\
& \hphantom{s.t. \:} w_k = \Delta_k (C_y x_k + D_y^u u_k) \\
& \hphantom{s.t. \:} [x_k^T, u_k^T]^T \in \mathbb{C}
\end{matrix*}
\label{eq_opt_1}
\end{equation}%
where $ \mathbb{C} = \{ [x^T, u^T]^T \mid H_x x + H_u u \le g \} $ defines the feasible set of states and control inputs.
\label{def_mmrmpc}
\end{definition}

The formulation of Definition \ref{def_mmrmpc} provides an exact solution to the optimal robust controller synthesis. However, such formulation is intractable due to its infinite dimension and due to the non-convexity of the optimization. Therefore, conservative approximations can be performed to address these two issues.

\subsection{Guaranteed Cost Control}

The first approximation is formulating a limit horizon problem that provides sufficient conditions to Definition 1. Towards that purpose, we define the guaranteed cost control approach proposed by \cite{xie1993control}.

\begin{definition}
A state feedback controller $ u_k = - K x_k $ is said to be a stabilizing guaranteed cost controller for the uncertain system \eqref{eq_sys_model} if there exists a symmetric matrix $ P \in \Re^{n_x \times n_x}, P \succ 0 $ that satisfies%
\begin{multline}
A_{cl}(\Delta)^T P A_{cl}(\Delta) - P + \\ + Q + N K + K^T N^T + K^T R K \preceq 0
\end{multline}%
such that $ J_i(x_0, \mathbf{u}, N) \le \bar{J}_u^*(x_0) = x_0^T P x_0 $ for the closed loop system%
\begin{equation}
A_{cl} = A + \delta A(\Delta_k) - \left( B^u + \delta B^u(\Delta_k) \right) K
\end{equation}%
subject to all admissible uncertainties $ \Delta_k \in \mathbb{D} $.
\label{def_gcc}
\end{definition}

\begin{definition}
A guaranteed cost feedback controller $ u_k = - K x_k $, satisfying Definition \ref{def_gcc}, with associated cost matrix $ P $ is said to be optimal if $ \tr(P) $ is infimal.
\label{def_optimal_gcc}
\end{definition}

For a discussion and interpretation of Definition \ref{def_optimal_gcc}, we refer the reader to Section 4.3 of \cite{massera2018optimal}.

\begin{lemma}
A state feedback controller $ u_k = - K x_k $ is said to be of guaranteed cost, according to Definition \ref{def_gcc}, if and only if there exist $ X \succ 0 $ and $ \Upsilon_i \succeq 0 $ for $ i \in [1, s] $ such that%
\begin{equation}
\begin{bmatrix}
- \Upsilon_q & 0 & 0 & C_y X - D_y^u Y\\ 
\star & - I_{n_c} & 0 & C_c X - D_c^u Y\\ 
\star & \star & - X + B^w \Upsilon_p B^{wT} & A X - B^u Y\\ 
\star & \star & \star & - X\\
\end{bmatrix} \preceq 0
\label{eq_gcc_lmi}
\end{equation}%
where $ X = P^{-1} $, $ Y = K P^{-1} $, the cost matrices $ C_c \in \Re^{n_c \times n_x} $ and $ D_c^u \in \Re^{n_c \times n_u} $ are given by the factorization of the cost function matrices%
\begin{equation}
\begin{bmatrix}
Q & N\\ 
N^T & R
\end{bmatrix} = \begin{bmatrix}
C_c & D_c^u 
\end{bmatrix}^T \begin{bmatrix}
C_c & D_c^u 
\end{bmatrix}
\end{equation}%
and the generalized S-Procedure variables are given by%
\begin{equation}
\begin{aligned}
\Upsilon_p &= \text{diag}(\upsilon_1 I_{n_{p1}}, \upsilon_2 I_{n_{p2}}, \ldots, \upsilon_s I_{n_{ps}})\\
\Upsilon_q &= \text{diag}(\upsilon_1 I_{n_{q1}}, \upsilon_2 I_{n_{q2}}, \ldots, \upsilon_s I_{n_{qs}}).
\end{aligned}
\end{equation}

\begin{proof}
Refer to Lemma 4.5 of \cite{massera2018optimal}.
\end{proof}
\label{lem_gcc}
\end{lemma}

\begin{theorem}
A state feedback controller $ u_k = - K x_k $ is said to be an optimal guaranteed cost controller, according to Definition \ref{def_optimal_gcc}, if and only if%
\begin{equation}
\begin{aligned}
K = \argmin\limits_{K} \;\; & \tr(Z) \\
s.t. \;\; & \begin{bmatrix}
-Z & I_{n_x}\\
\star & -X
\end{bmatrix} \preceq 0\\
& \text{Condition \eqref{eq_gcc_lmi}}
\end{aligned}
\label{eq_optimal_gcc}
\end{equation}%
where $ Z \in \Re^{n_x \times n_x} $ is the relaxation of the cost matrix $ P $ .

\begin{proof}
Refer to Theorem 4.8 and Remark 4 of \cite{massera2018optimal}.
\end{proof}
\label{the_optimal_gcc}
\end{theorem}

\subsection{Robust Positive Invariant Sets}

We now present the robust positive invariant (RPI) set definition.

\begin{definition}
The set $ \mathbb{R} \subseteq \mathbb{C} $ is said to be RPI for an autonomous uncertain linear system governed by $ x_{k+1} = (A + \delta A(\Delta)) x_k $ if and only if%
\begin{equation}
\forall x \in \mathbb{R}, \forall \Delta \in \mathbb{D}: (A + \delta A(\Delta)) x \in \mathbb{R}
\end{equation}%
or equivalently,%
\begin{equation}
\forall \Delta \in \mathbb{D}: (A + \delta A(\Delta)) \mathbb{R} \subseteq \mathbb{R}.
\end{equation}
\label{def_rpi}
\end{definition}

For diagonal disturbances, we can enumerate the vertexes of the disturbance set.
However, we cannot enumerate them for non-diagonal disturbances. Therefore, polytopic approaches are not applicable to the general case. Ellipsoidal RPI sets were widely studied for such cases (e.g. \cite{blanchini1999set} and \cite{kolmanovsky1998theory}). 
In this paper, they are employed to guarantee the infinite time constraint for a finite horizon controller. The separability of the uncertain dynamics is also  guaranteed, according to  \cite{rakovic2012homothetic}. Towards this purpose, the next definition presents the minimum RPI.

\begin{definition}
Let $ \Gamma $ be a set that satisfies Definition \ref{def_rpi}. Then, the set $ \mathbb{R}^{min} \subseteq \mathbb{C} $ is said to be minimal RPI if and only if $ \mathbb{R}^{min} \in \Gamma $ and $ \forall \mathbb{R} \in \Gamma: \mathbb{R}^{min} \subseteq \mathbb{R} $.
\label{def_min_rpi}
\end{definition}

\subsection{Principle of Separable Control Policies}

The principle of separable control policies has been employed in all tube-based model predictive controls, from rigid TMPC \cite{langson2004robust}, to parametrized TMPC \cite{rakovic2012parameterized}. It is one of the main contributions that enable such methods to be less conservative. The separated dynamics considered in this paper are defined as%
\begin{equation}
\begin{matrix*}[l]
z_{k+1} &= A z_k + B^u \nu_k \\
e_{k+1} &= A e_k + B^u \rho_k + B^w w_k \\
y_k &= C_y (z_k + e_k) + D_y^u (\nu_k + \rho_k)
\end{matrix*}
\label{eq_sys_model_separated}
\end{equation}%
where $ z_k $ defines the nominal dynamics and $ e_k $ defines the error dynamics. Eqs. \eqref{eq_sys_model_separated} are equivalent to the System \eqref{eq_sys_model} under $ x_k = z_k + e_k $ and $ u_k = \nu_k + \rho_k $.

\section{ROBUST CONTROL INVARIANT LEVEL-SETS}
\label{sec_mrci}

In this section, we propose a novel robust positive invariant (RPI) level-set formulation for linear systems subject to multiplicative uncertainties. The proposed $\sigma$-parameterized level-set $ \mathbb{R}(\sigma) $ is a direct generalization of Definition \ref{def_rpi} which provides invariance properties for disturbances level-sets $ \mathbb{W}(\sigma) $. It is valid only within an admissible state level-set $ \mathbb{Y}(\sigma) $.

\subsection{RPI Level-Set Formulation}

\begin{lemma}
    Consider the System $ x_{k+1} = A x_k + B^w w_k $. Then, $ \mathbb{R}(\sigma) $ is a RPI level-set if the following properties hold:%
    \begin{enumerate}
        \item[(i)] $ \forall \sigma \ge 0: \mathbb{R}(\sigma) \subseteq \mathbb{Y}(\sigma) $
        \item[(ii)] $ \forall \sigma \ge 0: A \mathbb{R}(\sigma) \oplus B^w \mathbb{W}(\sigma) \subseteq \mathbb{R}(\sigma) $ where $ \oplus $ is the Minkowski sum of sets
    \end{enumerate}%
    where%
    \begin{equation}
        \forall \sigma \ge 0: x_k \in \mathbb{Y}(\sigma) \Rightarrow w_k \in \mathbb{W}(\sigma).
    \end{equation}
    
    \begin{proof}
        To prove sufficiency assume $ \forall \sigma > 0\; \mathbb{R}(\sigma) $ is an RPI and \textit{(i)} holds. Then, $ x_k \in \mathbb{R}(\sigma) \Rightarrow w_k \in \mathbb{W}(\sigma) $, and%
        \begin{equation}
            \forall \sigma > 0, \forall x \in \mathbb{R}(\sigma), \forall w \in \mathbb{W}(\sigma): A x + B^w w \in \mathbb{R}(\sigma).
        \end{equation}%
        Therefore, \textit{(ii)} holds. Similarly, to prove necessity, if we assume \textit{(i)} doesn't hold, %
        \begin{equation}
            \exists x \in \mathbb{R}(\sigma), \exists w \not\in \mathbb{W}(\sigma): A x + B^w w \not\in \mathbb{R}(\sigma).
        \end{equation}%
        Therefore, \textit{(ii)} doesn't hold and $ \mathbb{R}(\sigma) $ isn't an RPI.
    \end{proof}
    \label{lem_rpi}
\end{lemma}

In the context of this paper, we are interested in the special case where the level sets are ellipsoidal. Therefore, we consider the case%
\begin{align}
    \mathbb{Y}(\sigma) &= \{ x \mid x^T E_Y x \le \sigma^2 \}
    \label{eq_ellipsoidal_y}\\
    \mathbb{W}(\sigma) &= \{ w \mid w^T E_W w \le \sigma^2 \}  \;\; \mbox{and}
    \label{eq_ellipsoidal_w}\\
    \mathbb{R}(\sigma) &= \{ x \mid x^T E_R x \le \sigma^2 \}
    \label{eq_ellipsoidal_r}
\end{align}%
where $ E_Y \succ 0 $, $ E_W \succ 0 $, $ E_R \succ E_Y \succ 0 $.

\subsection{Dynamics and properties of time-varying RPI level-sets}

If we consider the parametric variable $\sigma$ to be time-varying, we can express sufficient conditions for its dynamics as a second-order cone for ellipsoidal sets.

\begin{lemma}
    Let $ \mathbb{R}(\sigma) $ be an RPI level-set satisfying Lemma \ref{lem_rpi}, and assume there exist $ \alpha_k > 0 $ and $ \sigma_k > 0$ such that $ x_k \in \mathbb{R}(\alpha_k)$, and $ x_k \in \mathbb{Y}(\sigma_k) $. Then, the dynamics of $ A \mathbb{R}(\alpha_k) \oplus B^w \mathbb{W}(\sigma_k) \subseteq \mathbb{R}(\alpha_{k+1}) $ can be simplified to%
    \begin{equation}
        \alpha_{k+1}^2 = a_\alpha \alpha_k^2 + a_\sigma \sigma_k^2
    \end{equation}%
    or, equivalently,%
    \begin{equation}
        \alpha_{k+1} \ge || \sqrt{a_\alpha} \alpha_k, \sqrt{a_\sigma} \sigma_k ||
    \end{equation}%
    if $ a_\alpha $ and $ a_\sigma $ satisfy the following property:%
    \begin{equation}
        \begin{bmatrix}
            - E_R & E_R A & E_R B^w & 0 \\
            \star & - a_\alpha E_R & 0 & 0 \\
            \star & \star & - a_\sigma E_W & 0 \\
            \star & \star & \star & a_\alpha + a_\sigma - 1
        \end{bmatrix} \preceq 0.
        \label{eq_rpi_dyn_1}
    \end{equation}
    
    \begin{proof}
        From the RPI property of $ \mathbb{R}(\alpha_{k+1}) $%
        \begin{equation}
            \forall x \in \mathbb{R}(\alpha_k), \forall w \in \mathbb{W}(\sigma_k): A x + B^w w \in \mathbb{R}(\alpha_{k+1})
        \end{equation}%
        which is equivalent to%
        \begin{eqnarray}
            \forall x^T E_R x \le \alpha_k^2 \nonumber \\ \forall w^T E_W w \le \sigma_k^2: 
            \begin{bmatrix} 
                A^T E_R A & A^T E_R B^{wT} \\
                \star & B^{wT} E_R B^{wT}
            \end{bmatrix} \le \alpha_{k+1}^2.
            \label{eq_rpi_dyn_proof_1}
        \end{eqnarray}

        From Lemma \ref{lem_rpi}, \eqref{eq_rpi_dyn_proof_1} must hold for $ \alpha_k \equiv \sigma_k \equiv \alpha_{k+1} \equiv \sigma $. Then, using such property and applying the $S$-Procedure \cite{iwasaki2000generalized} to \eqref{eq_rpi_dyn_proof_1}, we obtain%
        \scriptsize
        \begin{equation}
            \begin{bmatrix} 
                A^T E_R A - a_\alpha E_R & A^T E_R B^{wT} & 0 \\
                \star & B^{wT} E_R B^{wT} - a_\sigma E_W & 0 \\
                \star & \star & a_\alpha + a_\sigma - 1
            \end{bmatrix} \preceq 0
        \end{equation}%
        \normalsize
        which is equivalent to \eqref{eq_rpi_dyn_1} by applying the Schur complement to $ E_R $ terms, and applying the symmetric transform $ T = \text{diag}(I, I, I, \sigma^{-2}) $.
    \end{proof}
    \label{lem_rpi_dynamics}
\end{lemma}

\begin{corollary}
    Let the state and disturbance level-sets $ \mathbb{Y}(\sigma) $ and $ \mathbb{W}(\sigma) $ are instead defined by the intersection of ellipsoids as%
    \begin{align}
        \mathbb{Y}(\sigma) &= \left\{ x \left| \forall i \in [1, s]: x^T E_{Y,i} x \le \sigma_i \right. \right\} \\
        \mathbb{W}(\sigma) &= \left\{ w \left| \forall i \in [1, s]: w_i^T E_{W,i} w_i \le \sigma_i \right. \right\}
    \end{align}%
    where $ \sigma \in \Re^{s}, \sigma > 0$. Then, based on the S-Procedure properties, the dynamics of Lemma \ref{lem_rpi_dynamics} directly generalizes to %
    \begin{equation}
        \alpha_{k+1} \ge || \sqrt{a_\alpha} \alpha_k^2, \sqrt{a_{\sigma,1}} \sigma_{k,1}, \dots, a_{\sigma,s} \sigma_{k,s}^2 ||
    \end{equation}%
    and the existence condition \eqref{eq_rpi_dyn_1} generalizes to%
    \begin{equation}
        \begin{bmatrix}
            - E_R & E_R A & E_R B^w & 0 \\
            \star & - a_\alpha E_R & 0 & 0 \\
            \star & \star & - A_\Sigma E_W & 0 \\
            \star & \star & \star & - 1 + a_\alpha + \sum_{i = 1}^{s} a_{\sigma_i}
        \end{bmatrix} \preceq 0
        \label{eq_rpi_dyn_structured_1}
    \end{equation}%
    where $ A_\Sigma = \text{diag}(a_{\sigma_1} I_{n_{p1}}, \dots, a_{\sigma_s} I_{n_{ps}}) $ and $ E_W = diag(E_{W,1}, \dots, E_{W,s}) $.
\begin{proof}
Omitted for brevity.
\end{proof}
\label{cor_rpi_dynamics_structured}
\end{corollary}

\begin{corollary}
Consider the minimal $ \alpha_k $ given by $ \alpha_k^* = \min\limits_{x_k \in \mathbb{R}(\alpha_k^2)} \alpha_k $ and assume $ \sigma_k \equiv \sigma $. Then, $ \lim\limits_{k \rightarrow \infty} \alpha_k^* \le \sigma $.
\label{cor_minimal_sigma}
\end{corollary}

\begin{corollary}
Let $ \mathbb{R}(\alpha_k) \subseteq \epsilon^2 \mathbb{Y}(\sigma_k) $ for some $ \epsilon \in [0, 1) $ and consider the minimal $ \alpha_k $ and $ \sigma_k $. Then, the following properties hold:%
\begin{equation}
    \alpha_{k+1}^* \le \epsilon \alpha_k^*,\;\;\;
    || \sigma_{k+1}^* ||_2 \le \epsilon \alpha_k^*,\;\;\;
    \lim\limits_{k \rightarrow \infty} \alpha_k^* = 0.
\end{equation}
\end{corollary}

\subsection{Synthesis of minimal RPI level-sets}

 In this section, we propose a minimal RPI synthesis and its sub-optimal version. The first, directly minimizes $ \mathbb{R}(\sigma) $ volume but it is unable to simultaneously synthesize a stabilizing controller. The second, an approximate minimal RPI synthesis allows the simultaneous synthesis of a stabilizing controller but does not directly minimize $ \mathbb{R}(\sigma) $ volume.


\begin{theorem} [Minimal RPI]
$ \mathbb{R}(\sigma) $ is an ellipsoidal minimal Robust Positive Invariant level set for the System \eqref{eq_sys_model} subject to a given feedback controller $ u_k = - K x_k $ if there exists $ a_\alpha \in [0, 1] $ where the following SDP has a solution%
\begin{align}
\inf\limits_{E_R, a_\alpha, a_\sigma} & - \logdet (E_R)
\label{eq_ellipsoid_mrpi_cost} \\
s.t. & \begin{bmatrix}
- E_R & E_R \bar{A} & E_R B^w\\ 
\star & - E_R a_\alpha & 0\\ 
\star & \star & - A_\Sigma E_W
\end{bmatrix} \preceq 0
\label{eq_ellipsoid_mrpi_cond} \\
& a_\alpha + \sum\limits_{i = 1}^{s} a_{\sigma_i} \le 1
\label{eq_ellipsoid_mrpi_cond_2} \\
& \forall i \in {[1, s]}: \bar{E}_{Y,i} - E_R \preceq 0 
\label{eq_ellipsoid_mrpi_cond_3}
\end{align}%
where $ \bar{A} = A - B^u K $, $ \bar{E}_{Y,i} = E_{Y,i}^x - E_{Y,i}^u K $ and other variables satisfy Corollary \ref{cor_rpi_dynamics_structured}.

\begin{proof}
Conditions \eqref{eq_ellipsoid_mrpi_cond} and \eqref{eq_ellipsoid_mrpi_cond_2} ensure the solution satisfies property \textit{(ii)} of Lemma \ref{lem_rpi} by the application of \eqref{eq_rpi_dyn_structured_1}. Meanwhile, condition \eqref{eq_ellipsoid_mrpi_cond_3} satisfies \textit{(i)}. Finally, The cost function \eqref{eq_ellipsoid_mrpi_cost} minimizes the log of the determinant of $ E_R $, which is proportional to the volume of the ellipsoid level-set $ \mathbb{R}(\sigma) $.
\end{proof}
\label{the_mrpi}
\end{theorem}

\begin{theorem} [Approximate Minimal RPI]
$ \mathbb{R}(\sigma) $ is an ellipsoidal approximate minimal RPI level set for the System \eqref{eq_sys_model} subject to a given feedback controller $ u_k = - K x_k $ if there exists $ a_\alpha \in [0, 1] $ where the following SDP has a solution%
\begin{align}
\inf\limits_{X, Y, a_\alpha, a_\sigma} & \tr(X)
\label{eq_approx_mrpi_cost} \\
s.t. & \begin{bmatrix}
- X & A X - B^u Y & B^w\\ 
\star & - X a_\alpha & 0\\ 
\star & \star & - A_\Sigma E_W
\end{bmatrix} \preceq 0
\label{eq_approx_mrpi_cond} \\
& \forall i \in {[1, s]}: \begin{bmatrix}
- I_{n_{qi}} & E_{Y,i}^x X - E_{Y,i}^u Y\\ 
\star & - X
\end{bmatrix} \preceq 0 
\label{eq_approx_mrpi_cond_3}\\
& a_\alpha + \sum\limits_{i = 1}^{s} a_{\sigma_i} \le 1
\label{eq_approx_mrpi_cond_2}
\end{align}%
where $ E_R = X^{-1}$, $ Y = K X $, and other variables satisfy Corollary \ref{cor_rpi_dynamics_structured}.

\begin{proof}
From the results of Theorem \ref{the_mrpi}, we first apply the symmetric transform $ T = \text{diag}(X, X, I) $ to \eqref{eq_ellipsoid_mrpi_cond}, expand the closed loop system $ \bar{A} $, and replace $ Y = K X $ to obtain \eqref{eq_approx_mrpi_cond}. Similarly, on \eqref{eq_ellipsoid_mrpi_cond_3} we apply the Schur complement and the similarity transform $ T = \text{diag}(X, I) $ and replace $ Y = K X $ to obtain \eqref{eq_approx_mrpi_cond_3}. Finally, since the minimization of $ logdet(X) $ is concave, we replace it by a proxy metric to the ellipsoid volume that still yields a convex minimization, $ tr(X) $.
\end{proof}
\label{the_approx_mrpi}
\end{theorem}

\subsection{Apllication to Norm-bounded Uncertainties}

From the disturbance feedback inequality of System \eqref{eq_sys_model}, we have%
\begin{equation}
\forall i \in [1, s]: w_i^T w_i \le x^T \bar{C}_{y,i}^T \bar{C}_{y,i} x.
\label{eq_uncertainty_inequality}
\end{equation}%
where $ \bar{C}_{y,i} = (C_y - D_y^u K)_i $ for a given controller $ u_k = - K x_k $. If the level-sets $ \mathbb{Y}(\sigma) $ and $ \mathbb{W}(\sigma) $ are defined such that%
\begin{equation}
    E_{W,i} = I_{n_{pi}}\;\;\;E_{Y,i}^x = C_{y,i}\;\;\;E_{Y,i}^u = D_{y,i}^u
\end{equation}%
from Lemma 4.2 of \cite{massera2018optimal} we obtain%
\begin{equation}
    \forall \sigma > 0: x_k \in \mathbb{Y}(\sigma) \Rightarrow w_k \in \mathbb{W}(\sigma)
\end{equation}%
as a sufficient condition to \eqref{eq_uncertainty_inequality}. From the losslessness property \cite{massera2018optimal} of \eqref{eq_uncertainty_inequality}, such condition is also necessary if $\sigma$ is minimal based on Corollary \ref{cor_minimal_sigma}. Therefore,%
\begin{equation}
\mathbb{W}(\sigma) = \bigcup\limits_{\Delta \in \mathbb{D}} \Delta \bar{C}_y \mathbb{Y}(\sigma).
\end{equation}

\begin{remark}
The $\sigma$-level-sets $ \mathbb{Y}(\sigma) $ and $ \mathbb{W}(\sigma) $ can be interpreted as an additive disturbance model tight conservative approximations of the system's multiplicative uncertainty \eqref{eq_uncertainty_inequality}. It is valid within a particular domain of the state space.
\end{remark}

\section{TUBE GUARANTEED COST MODEL PREDICTIVE CONTROL}
\label{sec_tgcmpc}

In this section, we present the formulation of the tube-based guaranteed cost model predictive controller. Towards that purpose, we first address the infinite dimensionality of Definition \ref{def_mmrmpc}.

\begin{lemma}
Consider a controller $ u_k = - K x_k $ with an associated cost matrix $ P $, such that $ K $ and $ P $ satisfy Lemma \ref{lem_gcc} and let $ \mathbb{R}_N = \{ x \mid x^T E_N x \le 1 \} \subseteq \mathbb{C} $ be an arbitrary RPI set for the System \eqref{eq_sys_model} subject to controller $ K $. Then, the existence of a solution for%
\begin{equation}
\begin{matrix*}[l]
J_f^*(x_0) = & \underset{\mathbf{u}}{\inf} \; \underset{\mathbf{\Delta}}{\sup} & J_0(x_0, \mathbf{u}) + x_N^T P x_N\\
& s.t. & x_{k+1} = A x_k + B^u u_k + B^w w_k\\
& & w_k = \Delta_k (C_y x_k + D_y^u u_k) \\
& & [x_k^T, u_k^T]^T \in \mathbb{C} \\
& & x_N \in \mathbb{R}_N 
\end{matrix*}
\label{eq_opt_2}
\end{equation}%
is a sufficient condition to the existence of a solution for Definition \ref{def_mmrmpc}, and $ J_f^*(x_0) \ge J^*(x_0) $.

\begin{proof}
Assume $ \forall k \ge N: u_k = - K x_k $. Then, $ x_N \in \mathbb{R}_N \Rightarrow x_k \in \mathbb{R}_N \subseteq \mathbb{C} $. Therefore the feasibility of \eqref{eq_opt_2} implies the feasibility of \eqref{eq_opt_1}.

From Definition \ref{def_gcc}, the cost matrix $ P $ upper-bounds the cost function of the closed loop system for all admissible disturbances $ \Delta $. Therefore, following from the previous assumption, we obtain%
\begin{equation}
x_N^T P x_N \ge \lim\limits_{M \rightarrow \infty} J_N(x_N, - K \mathbf{x})
\end{equation}%
and $ J_f^*(x_0) \ge J^*(x_0) $.
\end{proof}
\label{lem_finite_mmmpc}
\end{lemma}

Another important result is the transformation of the cost function to a deviation from the GCC solution.%
\begin{lemma}
Let $ \forall k < N: u_k = - K x_k + v_k $, where $ K $ satisfies Lemma \ref{lem_gcc} with an associated cost matrix $ P $. Then,%
\begin{equation}
J_0(x_0, \mathbf{u}, N) + x_N^T P x_N = x_0^T P x_0 + \sum\limits_{k = 0}^{N-1} v_k^T \bar{R} v_k
\label{eq_cost_function_modified}
\end{equation}%
where%
\begin{equation}
\bar{R} = R + D_y^{uT} \Lambda_q D_y^u + B^u (P^{-1} - B^w \Lambda_p^{-1} B_{wT})^{-1} B^u.
\end{equation}

\begin{proof}
Direct generalization of (\cite{massera2017guaranteed} - Lemma 2).
\end{proof}
\label{lem_cost_function_equivalency}
\end{lemma}

We apply the principle of separable control policies with $ v_k = \nu_k + \rho_k $ to the closed-loop counterpart of \eqref{eq_sys_model}, and obtain the equivalent system%
\begin{equation}
\begin{matrix*}[l]
z_{k+1} &= \bar{A} z_k + B^u \nu_k \\
e_{k+1} &= \bar{A} e_k + B^u \rho_k + B^w w_k \\
y_k &= \bar{C}_y (z_k + e_k) + D_u^u (\nu_k + \rho_k).
\end{matrix*}
\label{eq_sys_model_separated_2}
\end{equation}

Based on System \eqref{eq_sys_model_separated_2}, the approximate minimal RPI level-set can be used to relax the error dynamics $ e_k $ conservatively.

\begin{lemma}
Consider the RPI level-set $ \mathbb{R}^{min}(\alpha_k^2) $ satisfying Theorems \ref{the_mrpi} or \ref{the_approx_mrpi} with associated variables $ \alpha_k $, $ \sigma_k $, and controller $ u_k = - K_R x_k $. Then, $ \forall k \ge 0: e_k \in \mathbb{R}^{min}(\alpha_k^2) $ holds for the System \eqref{eq_sys_model_separated_2} if the following properties hold:%
\begin{enumerate}
\item $ \bar{A}_R \mathbb{R}^{min}(\alpha_k^2) \oplus B^w \mathbb{W}(\sigma_k^2) \subseteq \mathbb{R}^{min}(\alpha_{k+1}^2) $, where $ \bar{A}_R = A - B^u K_R $
\item $ z_k \oplus \mathbb{R}^{min}(\alpha_k^2) \subseteq \mathbb{Y}(\sigma_k^2) $
\item $ \rho_k = - (K_R - K) e_k $.
\end{enumerate}

\begin{proof}
Assume $ \forall k \ge 0: e_k \in \mathbb{R}^{min}(\alpha_k^2) $, then from the System \eqref{eq_sys_model_separated_2}%
\begin{multline} 
\forall e_k \in \mathbb{R}^{min}(\alpha_k^2), \forall w_k \in \mathbb{W}(\sigma_k^2): \\ \bar{A} e_k + B^u \rho_k + B^w w_k \subseteq \mathbb{R}^{min}(\alpha_{k+1}^2)
\label{eq_lem_separability_proof_1}
\end{multline}%
where $ \sigma_k $ satisfies $ x_k = z_k + e_k \in \mathbb{Y}(\sigma_k^2) $. Then, \eqref{eq_lem_separability_proof_1} is equivalent to%
\begin{equation}
B^u \rho_k \oplus \bar{A} \mathbb{R}^{min}(\alpha_k^2) \oplus B^w \mathbb{W}(\sigma_k^2) \subseteq \mathbb{R}^{min}(\alpha_{k+1}^2).
\end{equation}%
From the definition of $ \mathbb{R}^{min}(\alpha_k^2) $, property \textit{(i)} holds if $ \rho_k $ satisfies property \textit{(iii)} and property \textit{(ii)} also holds.
\end{proof}
\label{lem_separability_mrpi}
\end{lemma}

\begin{remark}
Lemma \ref{lem_separability_mrpi} provides only sufficient conditions since there may exist a $ \rho_k \neq - (K_r - K) e_k $ for which property \textit{(i)} still holds.
\end{remark}

\begin{lemma}
The System \eqref{eq_sys_model_separated_2} can be conservatively approximated by%
\begin{align}
z_{k+1} &= \bar{A} z_k + B^u \nu_k 
\label{eq_sys_model_separated_3_1}\\
\alpha_{k+1} &\ge \left|\left|\left[
\sqrt{a_\alpha} \alpha_{k},
\sqrt{a_{\sigma_1}} (\sigma_k)_1,
\dots,
\sqrt{a_{\sigma_s}} (\sigma_k)_s
\right]\right|\right|_2 
\label{eq_sys_model_separated_3_2}\\
(\sigma_k)_i &\ge || (\bar{C}_y)_i z_k + (D_y^u)_i \nu_k || + || (C_y^\alpha)_i ||_2 \alpha_k
\label{eq_sys_model_separated_3_3}
\end{align}%
where%
\begin{equation}
(C_y^\alpha)_i = (C_y - D_y^u K_R)_i E_R^{-\frac{1}{2}}.
\end{equation}

\begin{proof}
Eq. \eqref{eq_sys_model_separated_3_1} follows directly from \eqref{eq_sys_model_separated_2}, and \eqref{eq_sys_model_separated_3_2} follows directly from Lemma \ref{lem_rpi_dynamics}. Therefore, we focus this proof on \eqref{eq_sys_model_separated_3_3}. Property \textit{(ii)} of Lemma \ref{lem_separability_mrpi} is equivalent to%
\begin{multline}
\forall i \in [1, n_s], \forall e_k \in \mathbb{R}(\alpha_k^2): \\ \begin{bmatrix}
\bullet
\end{bmatrix}^T
\begin{bmatrix}
(\bar{C}_y)_i^T (\bar{C}_y)_i & (\bar{C}_y)_i^T D_{y,i}^u\\ 
\star & D_{y,i}^{uT} D_{y,i}^u
\end{bmatrix} 
\begin{bmatrix}
z_k + e_k\\ 
\nu_k
\end{bmatrix} \le (\sigma_k)_i^2.
\label{eq_soc_system_proof_1}
\end{multline}%
From \eqref{eq_soc_system_proof_1}, we obtain%
\begin{align}
\left|\left| (\bar{C}_y)_i (z_k + e_k) + D_{y,i}^u u_k \right|\right|_2 &\le\\
\left|\left| (\bar{C}_y)_i z_k + D_{y,i}^u \nu_k \right|\right|_2 + \max\limits_{e_k \in \mathbb{R}(\alpha_k^2)} \left|\left| (\bar{C}_{Ry})_i e_k \right|\right|_2 &=\\ 
|| (\bar{C}_y)_i z_k + (D_y^u)_i \nu_k || + || (C_y^\alpha)_i ||_2 \alpha_k &\le (\sigma_k)_i 
\label{eq_soc_system_proof_2}
\end{align}%
where $ \bar{C}_{Ry} = C_y - D_y^u K_R $. Therefore, \eqref{eq_soc_system_proof_2} is a sufficient condition for \eqref{eq_soc_system_proof_1}.
\end{proof}
\label{lem_soc_system}
\end{lemma}

Lemma \ref{lem_soc_system} provides a conic conservative relaxation of Eq. \eqref{eq_sys_model} that does not depend on the uncertain variable $ \Delta_k $ and is compatible with a SOCP optimization formulation. However, Lemma \ref{lem_soc_system} also implies that $ v_k \in \nu_k \oplus (K - K_R) \mathbb{R}^{min}(\alpha_k^2) $. Therefore, we conservatively approximate Eq. \eqref{eq_cost_function_modified} by%
\begin{equation}
J_0(x_0, \mathbf{u}, N) + x_N^T P x_N \le x_0^T P x_o + \sum\limits_{k = 0}^{N - 1} \gamma_k^2
\label{eq_cost_relaxed}
\end{equation}%
where%
\begin{equation}
\gamma_k \ge || \bar{R}^\frac{1}{2} \nu_k ||_2 + || \bar{R}^\frac{1}{2} (K_R - K) E_R^{-\frac{1}{2}} ||_2 \alpha_k.
\end{equation}

The last remaining approximation requires to remove the uncertainty from the problem representation. It provides conservativeness to the controller feasible domain $ [x_k^T u_k^T]^T \in \mathbb{C} $ as a function of $ z_k $, $ \nu_k $ and $ \alpha_k $.

\begin{lemma}
Consider the system from Lemma \ref{lem_soc_system}. Then,%
\begin{multline}
[z_k^T, \nu_k^T, \alpha_k] \in \bar{\mathbb{C}} = \{ [x^T, \nu^T, \alpha^T]^T \mid \\ \forall i \in [1, n_c]: \bar{H}_i z + (H_u)_i \nu + || (\bar{H}_{R})_i X_R^\frac{1}{2} ||_2 \alpha \le g_i \}
\label{eq_robust_constraint}
\end{multline}%
where $ \bar{H} = H_x - H_u K $, and $ \bar{H}_R = H_x - H_u K_R $, implies $ [x_k^T, u_k^T]^T \subseteq \mathbb{C} $.

\begin{proof}
From the definition of $ \mathbb{C} $, we obtain%
\begin{equation}
H_x x_k + H_u u_k \le g.
\label{eq_constraint_proof_1}
\end{equation}%
From the separated dynamics Eqs. \eqref{eq_sys_model_separated_2} and Lemma \ref{lem_separability_mrpi}, \eqref{eq_constraint_proof_1} is equivalent to%
\begin{equation}
\forall e_k \in \mathbb{R}^{min}(\alpha_k^2): \bar{H} z_k + H_u \nu_k + \bar{H}_R e_k \le g.
\label{eq_constraint_proof_2}
\end{equation}%
Then, the robust inequality constraint counterpart for the $i$-th row of \eqref{eq_constraint_proof_2}, is given by%
\begin{equation}
\bar{H}_i z_k + (H_u)_i \nu_k + \max\limits_{e_k^T X^{-1} e_k \le \alpha_k} (\bar{H}_R)_i e_k \le g
\end{equation}%
which is equivalent to \eqref{eq_robust_constraint}.
\end{proof}
\label{lem_robust_constraint}
\end{lemma}

We are now ready to present the proposed T-GCMPC formulation.

\begin{theorem}
Consider the system from Lemma \ref{lem_soc_system}, subject to the guaranteed cost controller $ K $ with associated cost matrix $ P $ and a mRPI level-set $ \mathbb{R}^{min}(\alpha_k^2) $ with associated controller $ K_R $. Then, the optimization $ \bar{J}^*(x_0) $, given by%
\begin{equation}
\begin{matrix*}[l]
& \underset{\mathbf{\nu}}{\inf} & x_0^T P x_0 + \sum\limits_{k = 0}^{N - 1} \gamma_k^2\\
& s.t. & z_{k+1} = (A - B^u K) z_k + B^u \nu_k\\
& & \alpha_{k+1} \ge \left|\left|\left[
\sqrt{a_\alpha} \alpha_{k}, \sqrt{a_{\sigma_1}} (\sigma_k)_1, \dots, \sqrt{a_{\sigma_s}} (\sigma_k)_s \right]\right|\right|_2 \\ 
& & (\sigma_k)_i \ge || (\bar{C}_y)_i z_k + (D_y^u)_i \nu_k || + (C_y^\alpha)_i \alpha_k\\
& & \gamma_k \ge || \bar{R}^\frac{1}{2} \nu_k ||_2 + || \bar{R}^\frac{1}{2} (K_R - K) E_R^{-\frac{1}{2}} ||_2 \alpha_k \\
& & \bar{H}_i z_k + (H_u)_i \nu_k + || (H_R)_i E_R^{-\frac{1}{2}} ||_2 \alpha_k \le g_i \\
& & || E_N^{\frac{1}{2}} z_N ||_2 + || E_N^{\frac{1}{2}} E_R^{-\frac{1}{2}} ||_2 \alpha_N \le 1
\end{matrix*}
\label{eq_opt_tgcmpc}
\end{equation}%
is a conservative approximation, both in cost and feasibility, to Lemma \ref{lem_finite_mmmpc} and, by consequence, Definition \ref{def_mmrmpc}.

\begin{proof}
The optimization problem \eqref{eq_opt_tgcmpc} follows directly from Lemmas \ref{lem_cost_function_equivalency}, \ref{lem_separability_mrpi}, \ref{lem_robust_constraint}, \ref{lem_soc_system}, and the relaxed cost \eqref{eq_cost_relaxed}.
\end{proof}
\label{the_tgcmpc}
\end{theorem}

Theorem \ref{the_tgcmpc} presents the T-GCMPC formulation. It yields an SOCP optimization which provides a conservative approximation to the general intractable robust model predictive control problem from Definition \ref{def_mmrmpc}.


\section{NUMERICAL EXAMPLES}
\label{sec_examples}

In this section, we present a comparative study among the T-GCMPC proposed (Theorem \ref{the_tgcmpc}), the OL-GCMPC presented in \cite{massera2017guaranteed}, and the HTMPC proposed by Rakovi{\'c} and Cheng \cite{rakovic2013homothetic}. We use the YALMIP Toolbox \cite{lofberg2004yalmip} and the Mosek solver \cite{mosek} for the modeling of the problem.

\begin{figure*}[ht]
\begin{center}
\includegraphics[width=16cm,height=3cm]{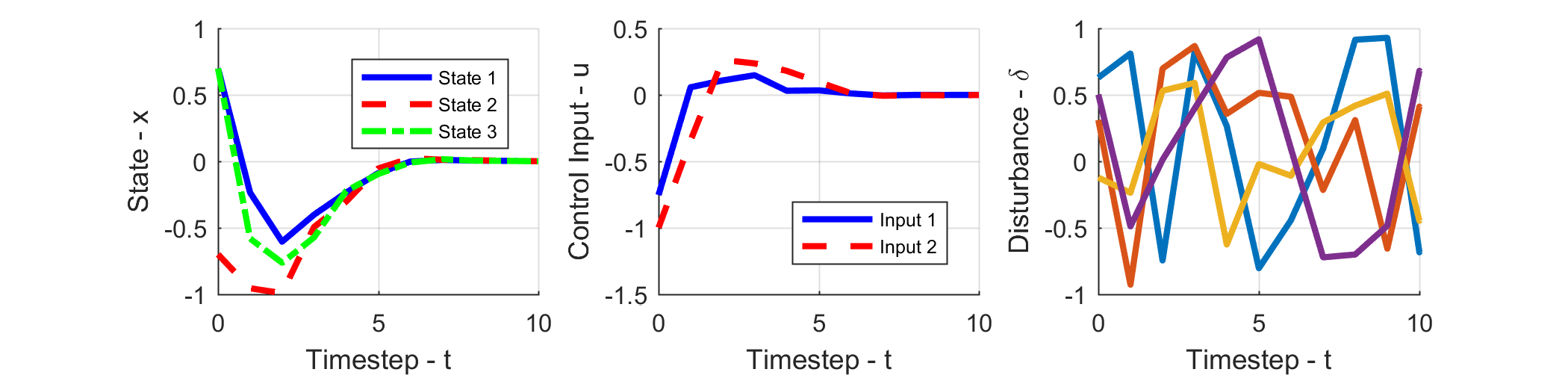}
(a)
\includegraphics[width=16cm,height=3cm]{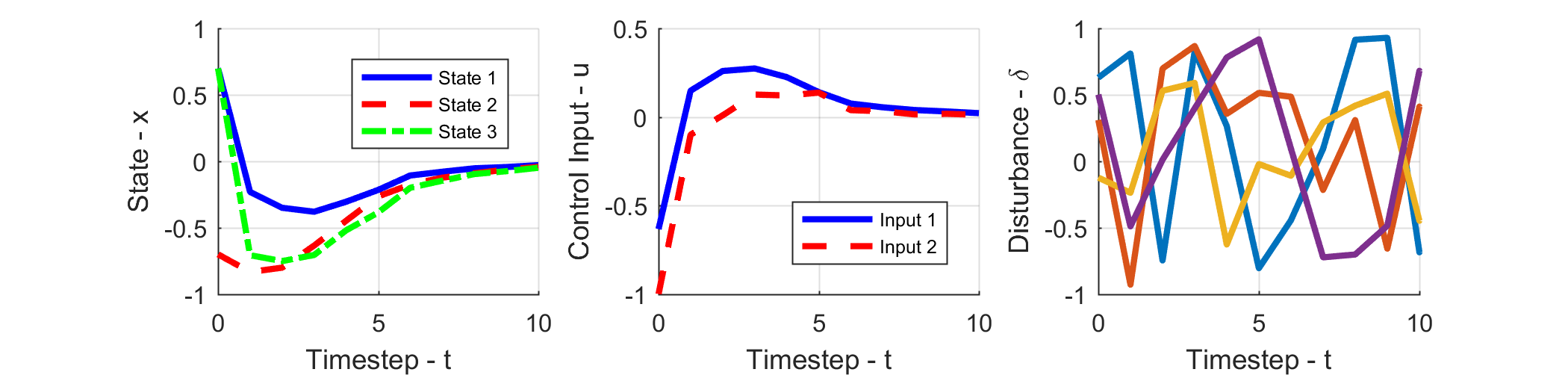}
(b)
\caption{State, control input and uncertainties for the system \eqref{eq_model_example} controlled by T-GCMPC (a) and HTMPC (b) approaches.}
\label{fig_tgcmpc_example}
\end{center}
\end{figure*}

\subsection{Investigated System}

Consider the uncertain linear system \eqref{eq_sys_model} defined by%
\begin{equation*}
\begin{matrix}
A = \begin{bmatrix}
1.1 & 0 & 0\\ 
0 & 0 & 1.2\\ 
-1 & 1 & 0
\end{bmatrix},
\;
B^u = \begin{bmatrix}
0 & 1\\ 
1 & 1\\ 
-1 & 0
\end{bmatrix},
\\
B^w = \begin{bmatrix}
0.17 & 0.07\\ 
0.12 & -0.1\\ 
-0.17 & 0.02
\end{bmatrix}
\\
C_y = \begin{bmatrix}
0.41 & 0.43 & -0.5\\ 
0 & -0.32 & 0.44
\end{bmatrix},
\;
D_y^u = \begin{bmatrix}
0.4 & -0.4\\ 
0 & 0
\end{bmatrix}
\end{matrix}
\label{eq_model_example}
\end{equation*}%
with a diagonal admissible uncertainty set ($ \Delta_k = \text{diag}(\Delta_{k1}, \Delta_{k2}) $), associated with weighting  matrices $ Q = I $, $ R = I $, and $ N = 0 $ and subject to the constraints%
\begin{equation}
H_x = \begin{bmatrix}
I_3\\ -I_3 \\ 0 \\ 0
\end{bmatrix},\; H_u = \begin{bmatrix}
0\\ 0 \\ I_2 \\ -I_2
\end{bmatrix},
\text{ and } 
g = \mathbbm{1}.
\end{equation}
 The GCC synthesis yields%
\begin{equation}
K = \begin{bmatrix}
0.05 & -0.27 & 0.46\\ 
1.89 & -0.55 & -0.43
\end{bmatrix}
\end{equation}%
with associated cost matrix and control input deviation cost%
\begin{equation}
P = \begin{bmatrix}
19.18 & -5.98 & -9.57\\ 
\star & 4.42 & 2.47\\
\star & \star & 7.21
\end{bmatrix},\;\;
%
%
\bar{R} = \begin{bmatrix}
10.92 & 10.80\\ 
\star & 22.85
\end{bmatrix}. \nonumber
\end{equation}
For simplifying purposes we assume $ K_R = K $, which yields an approximate mRPI level set defined by%
\begin{equation}
E_R = \begin{bmatrix}
1.06 & 1.44 & 1.21 \\ 
\star & 2.52 & 1.68\\ 
\star & \star & 1.67
\end{bmatrix}^{-1}
\end{equation}%
with associated cross-section variables%
\begin{equation}
\begin{matrix*}[l]
a_\alpha = 0.48, \;\;\;a_\sigma = \begin{bmatrix} 
0.34 & 0.17
\end{bmatrix}^T. 
\end{matrix*}
\end{equation}

\subsection{Controller Comparison}

Towards an unbiased comparison, we do not consider a terminal constraint set in this example. We modify the formulation of the HTMPC from \cite{rakovic2013homothetic} to incorporate an identical ellipsoidal tube for closer comparison. We choose a horizon $ N = 5 $ due to the limitations of the OL-GCMPC where increasingly larger horizons yield increasingly smaller feasible regions.

Figures \ref{fig_tgcmpc_example} presents  simulation results of the T-GCMPC and the HTMPC for the initial state $ x_0 = [0.7, -0.7, 0.7]^T $. This initial state is outside the feasible region for the OL-GCMPC. We can observe that both controllers can stabilize the system within the feasible state region $ x_i \in [-1, 1] $ and $ u_i \in [-1, 1] $ while subject to a structured norm bounded disturbance. It is worth noting that the HTMPC results in a larger settling time than the T-GCMPC. However, the tuning of both controllers is not directly comparable due to their different formulations, especially with regards to the penalization of the parameter $ \alpha $.

\begin{figure}[ht]
\centering
\includegraphics[width=8cm,height=3.7cm]{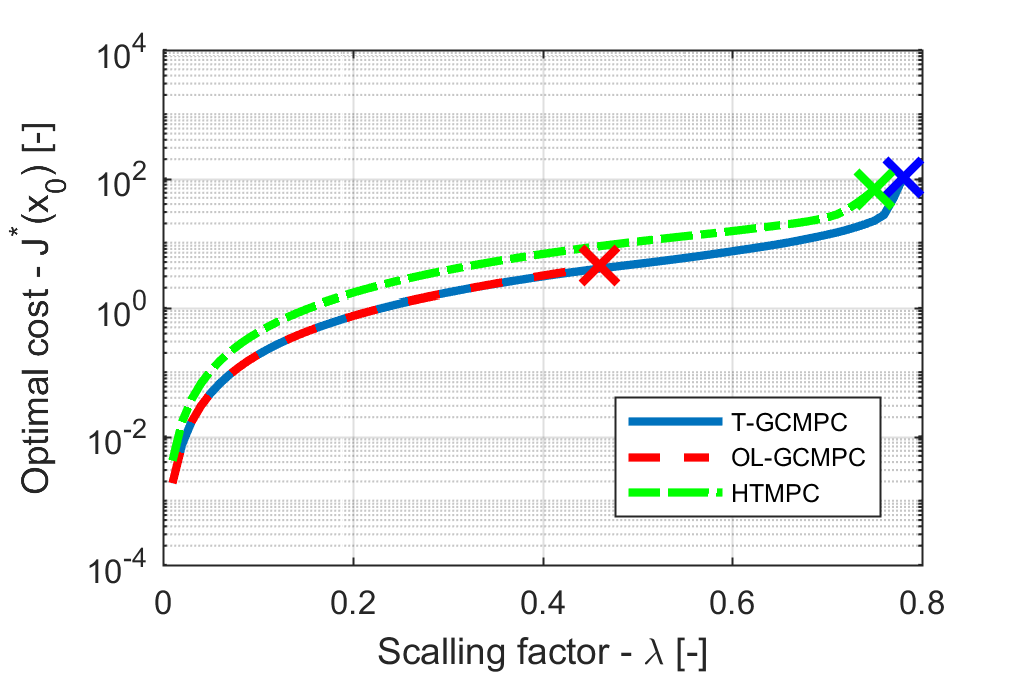}
\caption{Optimal cost of T-GCMPC (blue solid line) and OL-GCMPC (red dashed line).}
\label{fig_cost_gcmpc}
\end{figure}
Figure \ref{fig_cost_gcmpc} presents the resulting costs of the controllers for the manifold $ x_0 = \lambda [1, -1, 1]^T $ where $ \lambda \in [0, 0.8] $. It is possible to observe that the T-GCMPC yields a larger feasible domain $ \lambda \in [0, 0.78] $ when compared to the OL-GCMPC domain $ \lambda \in [0, 0.46] $ and to the HTMPC domain $ \lambda \in [0, 0.75] $. The crosses in Fig. \ref{fig_cost_gcmpc} indicate the largest value of $ \lambda $ where both controllers are feasible. It also ensures a lower or equal cost if compared to the HTMPC and the OL-GCMPC, respectively. Notice that longer horizons would yield an even larger difference among feasible regions of the controllers. This difference is due to the conservative nature of the OL-GCMPC with respect to the uncertainties propagation through the optimization horizon.

\begin{figure}[ht]
\centering
\includegraphics[width=8cm,height=3.5cm]{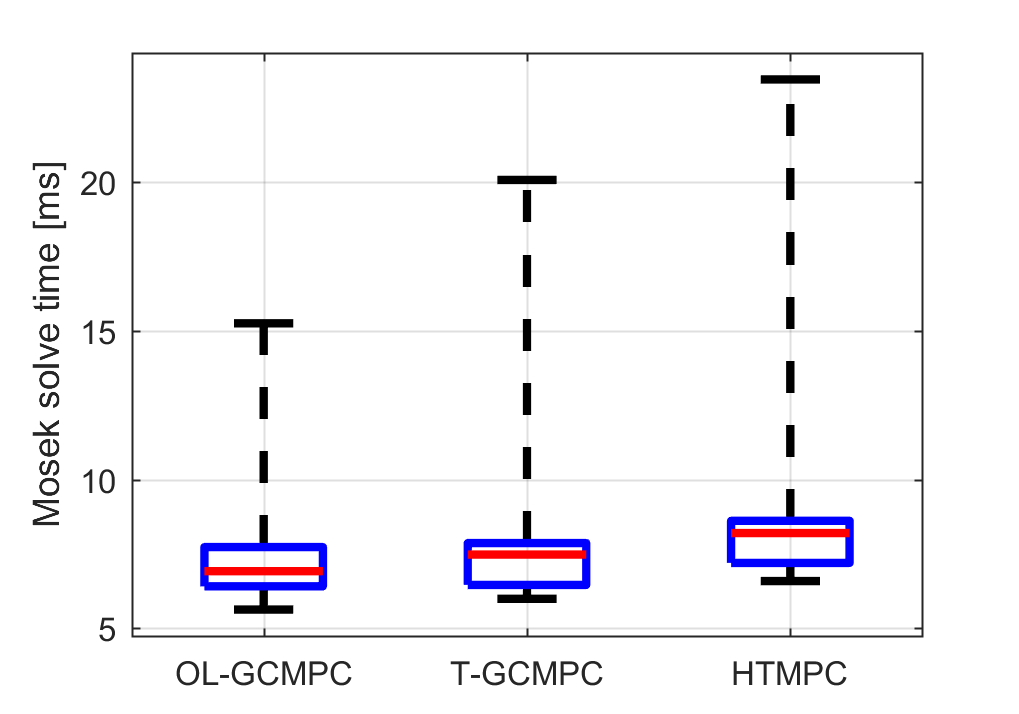}
\caption{Box-plot of solution time of $1000$ executions for both the OL-GCMPC, T-GCMPC and HTMPC.} 
\label{fig_time_boxplot}
\end{figure}
\subsection{Timing Analysis}

In order to obtain the solutions for  T-GCMPC, OL-GCMPC and HTMPC, are required minima  of $5.92ms$, $5.70 ms$ and $6.57ms$; maxima of $19.04ms$, $15.27 ms $ and  $22.44ms$; and averages  of $7.24ms$, $7.01ms$, and $7.90ms$, respectively. We obtain these results on a dual $2.6Ghz$ Intel Xeon E5-2670 with $32Gb$ of RAM. Figure \ref{fig_time_boxplot} presents the computational time distribution box plots, where it is possible to see that all controllers have comparable execution times. The T-GCMPC requires $ 9.3 \% $ longer execution time than the OL-GCMPC and $ 2.2 \% $ shorten execution time than the HTMPC on the $99^{th}$ percentile of the distribution. However, due to the vertex enumeration approach, the HTMPC requires $ 2^{n_p} $ constraints per timestep of the horizon to guarantee a robust dynamics. On the other hand, the T-GCMPC requires $ n_p + 2 $ constraints per timestep. It is more effective for  systems subject to a higher number of disturbance modes.

\subsection{T-GCMPC Optimization Results}

\begin{figure}[ht]
\centering
\includegraphics[width=8cm,height=4cm]{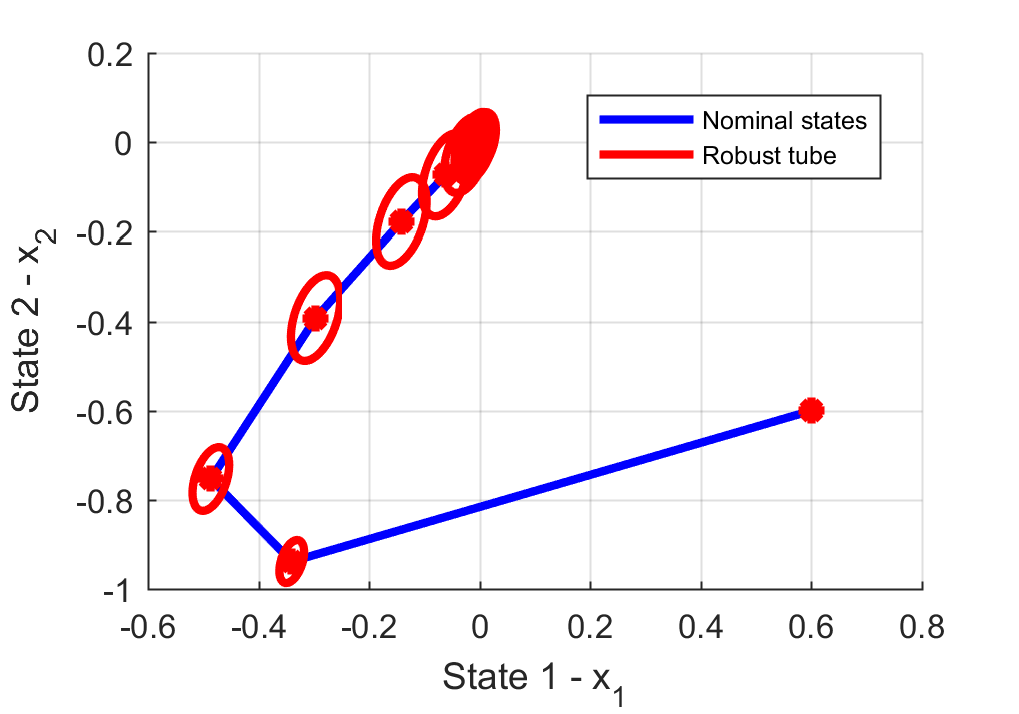}
\caption{T-GCMPC optimization  for states $ x_1$ and $ x_2 $.}
\label{fig_tube_example}
\end{figure}

To illustrate the result of the T-GCMPC optimization, Figure \ref{fig_tube_example} presents the nominal states (blue) and the invariant tube (red) for the initial state $ x_0 = [0.6, -0.6, 0.6]$ for a horizon $ N = 20 $. We can observe that the tube cross-section grows initially due to the magnitude of the disturbances. When the system approaches the origin, the tube cross-section tends to zero.



\section{CONCLUSION}
\label{sec_conclusion}

We  proposed a tube-based guaranteed cost model predictive controller based on homothetic invariant tube formulation for constrained linear systems,  subject to  norm-bounded uncertainties. The proposed method provides an upper bound to the general RMPC cost function. It also internalizes all the system uncertainties through its homothetic tube formulation, which guarantees that its cross-section volume tends to zero (i.e., all uncertainties are guaranteed to be rejected). We have also proposed a novel invariant tube definition, where any arbitrary scaling of its cross-section is an invariant set for a subset of the state-space. The method provides cost and feasibility guarantees while yielding a second order cone programming problem. In general it outperforms computationally similar methods solved based on semi-definite programming problems. Additionally, the T-GCMPC is less conservative than the OL-GCMPC while providing stronger robustness guarantees.



Future works include the study of polytope-represented mRPI to reduce the optimization complexity from SOCP to a QP or QCQP and the generalization of the proposed controller to reference-tracking problems.

\bstctlcite{BSTcontrol}
\bibliographystyle{IEEEtranS}
\bibliography{refs_paper}
\end{document}